\documentclass[runningheads]{llncs}
\usepackage{graphicx}
\usepackage{tabularx}
\usepackage{makecell}
\usepackage{pdflscape}

\usepackage{pdflscape}
\usepackage{array}

\usepackage{colortbl}

\newcolumntype{P}[1]{>{\raggedright\arraybackslash}p{#1}}

\definecolor{lightgray}{gray}{0.9}
\definecolor{white}{rgb}{1,1,1}

\newcommand{\eg}{e.\,g. }
\newcommand{\etal}{et\,al. }
\newcommand{\ie}{i.\,e. }

\newcommand{\doubleq}[1]{``#1''}

\begin{document}
\title{Teaching Quantum Informatics at School: Computer Science Principles and Standards}
\titlerunning{Quantum Informatics at School}
%
\author{Giulia Paparo\inst{1}\orcidID{0000-0002-4782-8337} \and Regina Finsterhoelzl \inst{2}\orcidID{0000-0002-0899-4957} \and
Bettina Waldvogel\inst{1}\orcidID{0000-0002-0658-1032} \and
Mareen Grillenberger\inst{1,3,4}\orcidID{0000-0002-8477-1464}}
\authorrunning{G. Paparo et al.}
%
\institute{Schwyz University of Teacher Education, Goldau, Switzerland \email{\{giulia.paparo,bettina.waldvogel,mareen.grillenberger\}@phsz.ch} \and
University of Konstanz, Konstanz, Germany
\email{regina.finsterhoelzl@uni-konstanz.de} \and
Lucerne University of Teacher Education, Lucerne, Switzerland \and
Lucerne University of Applied Sciences and Arts, Rotkreuz, Switzerland
}
\maketitle              
\begin{abstract}
The application of knowledge from quantum physics to computer science, which we call \doubleq{quantum informatics}, is driving the development of new technologies, such as quantum computing and quantum key distribution. Researchers in physics education have recognized the promise and significance of teaching quantum informatics in schools, and various teaching methods are being developed, researched and applied. Although quantum informatics is equally relevant to computer science education, little research has been done on how to teach it with a focus on computer science concepts and knowledge. In this study, we position quantum informatics within Denning's Great Principles of Computing and propose Quantum Informatics Standards for secondary schools. 

\end{abstract}

\section{Introduction}
The application of quantum physics to computer science has the potential to open new avenues of thought and endeavors within computer science. From quantum computers that could outperform classical supercomputers at solving computationally hard problems, to quantum key distribution that could make it physically impossible for a third party to eavesdrop unnoticed, quantum informatics has the potential to have a major impact on computer science and society as a whole \cite{de_wolf_potential_2017}.
Most of these technologies are still in development, and the timeframe and applications for their large-scale implementation are not yet predictable. However, this does not affect the theoretical change in thinking and skills that quantum technologies require. It is important to start thinking about how to introduce students to this new way of thinking at an early stage, for several reasons. First, teaching quantum informatics in schools means laying the foundation for an informed society that can consciously discuss the future of these technologies and their applications. Second, learning about quantum information can help students develop new perspectives, challenge what they think they already know, and develop a curious approach to the nature of computation. Finally, from an educational perspective, it is a wonderful example to explore how to teach complex and inherently multidisciplinary topics in school.

The aim of this paper is to support the teaching of quantum informatics in lower and upper secondary schools by providing an overview of the subject and defining principles and standards, as an orientation in the development of educational material. To this end, we ask the following questions: \textit{How can quantum informatics be positioned from the perspective of computer science education?} and \textit{What are the most important learning outcomes for secondary school
students learning about quantum informatics from a computer science education perspective?} To answer these questions, we first suggest how quantum informatics could be positioned within Peter Denning's Great Principles of Computing \cite{denning_great_nodate}, and then propose Quantum Informatics Standards for secondary schools.

\section{Related Works}
In the past years, there has been a rapidly growing interest in the teaching of quantum informatics. This interest sparked especially from physics educators, who saw in quantum technologies the possibility to access quantum mechanics, a notorious hard to teach subject at school, in a more tangible and direct way \cite{deutsch_physics_nodate}. Many interesting approaches for teaching quantum informatics at school were developed and evaluated, \eg \cite{perry_quantum_2020}, \cite{billig_quantum_2018}, \cite{economou_teaching_2020}. In addition, a variety of serious games \cite{seskir_quantum_nodate}, visualization tools \cite{migdal_visualizing_2022}, zines (\ie self-published DIY magazines) \cite{franklin_exploring_2020}, and role-playing \cite{lopez-incera_entangle_2019} approaches were developed that could be used to engage high school students along with the broader public. We will present in more depth some of these approaches in section \ref{teachingexamp}. It is important to note that these were mostly small-scale interventions carried out by a group of experts, and that more extensive research, as well as teaching guidelines, training, or evaluation criteria for teachers are still needed. 

An important step towards establishing a common language as well as an orientation for educational programs was the formulation of the European Framework for Quantum Technologies \cite{directorate-general_for_communications_networks_competence_2021}. However, this framework aims to map competencies and skills for quantum technologies, and therefore includes topics of little relevance from a computer science perspective, such as quantum sensing. At the same time, it does not give much space to relevant topics, such as quantum information theory, and focuses mainly on higher education. On the contrary, \doubleq{Key Concepts for Future Quantum Information Science Learners} \cite{alpert_key_nodate} and more directly its expansion for computer science \cite{franklin_qis_2021} target more clearly quantum informatics from a computer science perspective and at high school level. The first \cite{alpert_key_nodate} offers a good overview and a basis for educators, and it is expanded and evaluated further for Computer Science learning outcomes and activities \cite{franklin_qis_2021}. The latter, however, is aimed at pre-college and beyond, and the level of knowledge and skills required is not suitable to younger students. 

Seegerer \etal \cite{seegerer_quantum_2021} discussed quantum computing as a topic in computer science education and made a proposal for its central concepts, ideas, and explanatory approaches. In our previous work, we systematically analyzed the key concepts of quantum informatics further and applied them for the formulation of competencies within guidelines of the German Informatics Society \cite{paparo}. Although the results are in German, the presented mapping of the key concepts of quantum informatics should also be understandable for English speakers, as it consists mostly of language-independent technical terms. 

The above-mentioned works constitute a first good step for supporting the teaching of quantum informatics. Standards for quantum informatics education at the lower and upper secondary level are still needed, and this work aims to fill this gap, building upon some of the previously defined key concepts and competencies (\cite{franklin_qis_2021}, \cite{seegerer_quantum_2021}, \cite{paparo}, \cite{directorate-general_for_communications_networks_competence_2021}).

\section{Terminological Remarks}
Generally, it is distinguished between quantum technologies of the first generation and second generation. The first generation is defined by technologies that could be built thanks to our understanding of quantum physics. This definition includes technologies, such as lasers, transistors and global positioning systems (GPS). The second generation of quantum technologies builds on the capability to control and manipulate the properties of quantum systems, for example, to build quantum computers and quantum sensors. 

Naturally, as computer science educators, we are looking further than the technology alone. As Michaeli \etal \cite{michaeli_quanteninformatik_2021} pointed out, the best term to define the field of research describing the knowledge and applications related to these new technologies from the perspective of computer science is in German \doubleq{Quanteninformatik}. 
We translated that as \doubleq{quantum informatics} instead of quantum computer science to indicate a broader view on information processing not limited to computers alone. Other commonly used terms to describe the field are \doubleq{Quantum Information Science (QIS)}, \doubleq{Quantum Information Science and Technologies (QIST)}, and \doubleq{Quantum Computing}. All of these terms intersect in some aspects and at the same time suggest a different focus (\ie on information theory or computation). The term \doubleq{quantum informatics}, although less commonly used, describes more directly what we are interested in: We want students to understand, use, and think about the changes in computer science (informatics) brought about by the application of our knowledge of quantum theory. For these reasons, we prefer the term \doubleq{quantum informatics} in this article and hope it will be broadly used in the future. 

\section{Methodological approach}
In order to position quantum informatics from the perspective of computer science, we used the Great Principles framework of Peter Denning \cite{denning_great_nodate} as a way to categorize a new technology within overarching principles that could then serve as a guide for educators and curriculum developers. As described in more detail in section \ref{principles}, we first categorized the key concepts of quantum informatics \cite{paparo} within the Great Principles. The categorization was carried out by the first author, while the second and third authors, with expertise in quantum information theory and computer science education respectively, provided critical review. Re-categorization possibilities were discussed, as well as the addition of further concepts. The categorization was then reviewed by an external computer science expert.

The newly formulated orientation within the Great Principles, as well as relevant aspects of the previously described frameworks (\cite{franklin_qis_2021}, \cite{directorate-general_for_communications_networks_competence_2021}, \cite{seegerer_quantum_2021}), then served as a basis for the further formulation of the standards. The previously defined learning goals \cite{paparo} were analyzed within the framework of the CSTA K-12 Computer Science Standards \cite{computer_science_teacher_association_csta_nodate}. These describe a general core set of learning objectives for the K-12 computer science curriculum and are widely used for curriculum development and assessment in the USA and around the world, and thus provided an excellent orientation for an initial formulation of international quantum informatics standards. The formulated standards were iterated following the principles of the Standards Developer's Guide \cite{k12_computer_science_framework_guidance_nodate}. Details are explained in section \ref{standards}. Finally, to make the learning outcomes more tangible to the reader, exemplary teaching approaches from the literature were assigned to the standards.  The teaching approaches were chosen based on whether they were considered to fit the newly defined Quantum Informatics Standards, specifically whether they were designed for high school students, address one or more of the defined standards, and were suitable from a computer science perspective (see section \ref{teachingexamp} for more details). 
 
\section{Great Principles of Quantum Informatics} \label{principles}
Peter Denning defined in a compact and coherent way, the overarching, fundamental principles of computer science, what he called \doubleq{Great Principles of Computing} \cite{denning_great_nodate}. The Great Principles framework aimed to stimulate deep structural thinking about computer science as a discipline and to provide a common language that encourages connections within computer science and across disciplines, providing existing stories while inspiring and structuring new didactic approaches. To do so, he defined seven categories (\doubleq{windows}) based on the fundamental questions underlying computing technologies. For each window, Denning proposed \doubleq{principal stories} to depict the richness of each principle, make it more tangible, and to indicate examples of its historical development.

We carried out a categorization of quantum informatics within the Great Principles framework, in order to provide an orientation of quantum informatics within computer science, \ie to help in structuring and seeing the principles of computer science within quantum informatics. To do so, quantum informatics concepts \cite{paparo} were first categorized within this framework. The categorization was then critically reviewed, and re-categorization and the addition of new concepts were discussed. There was almost no need for re-categorization. As in Denning’s framework, some concepts could be assigned to more than one window, as these are regarded as overlapping rather than exclusive. However, in order to achieve a comparable balance with the principal stories of classical computing, it was necessary to add several concepts that were not originally part of our selection. This was especially the case for the windows \doubleq{Design} and \doubleq{Evaluation}, which were also added later to the Great Principles by Peter Denning \cite{denning_computing_2007}. As Denning states, this framework is evolving with time and is not an exhaustive representation of the field, which is even more true for a relatively young discipline such as quantum informatics. 

Our goal was to use the framework as a didactic tool to provide orientation, not to revise the Great Principles. While it has been possible to position the main concepts of quantum informatics within this framework, this should not be seen as an exhaustive or conclusive statement, but rather as an exploratory approach to provide guidance to computer science teachers and educators. The results are illustrated in Table \ref{fig:denning}.

\begin{table}\caption{Great Principles of Computing and Principal-Stories of Quantum Informatics} \label{fig:denning}   \includegraphics[width=1
 \linewidth]{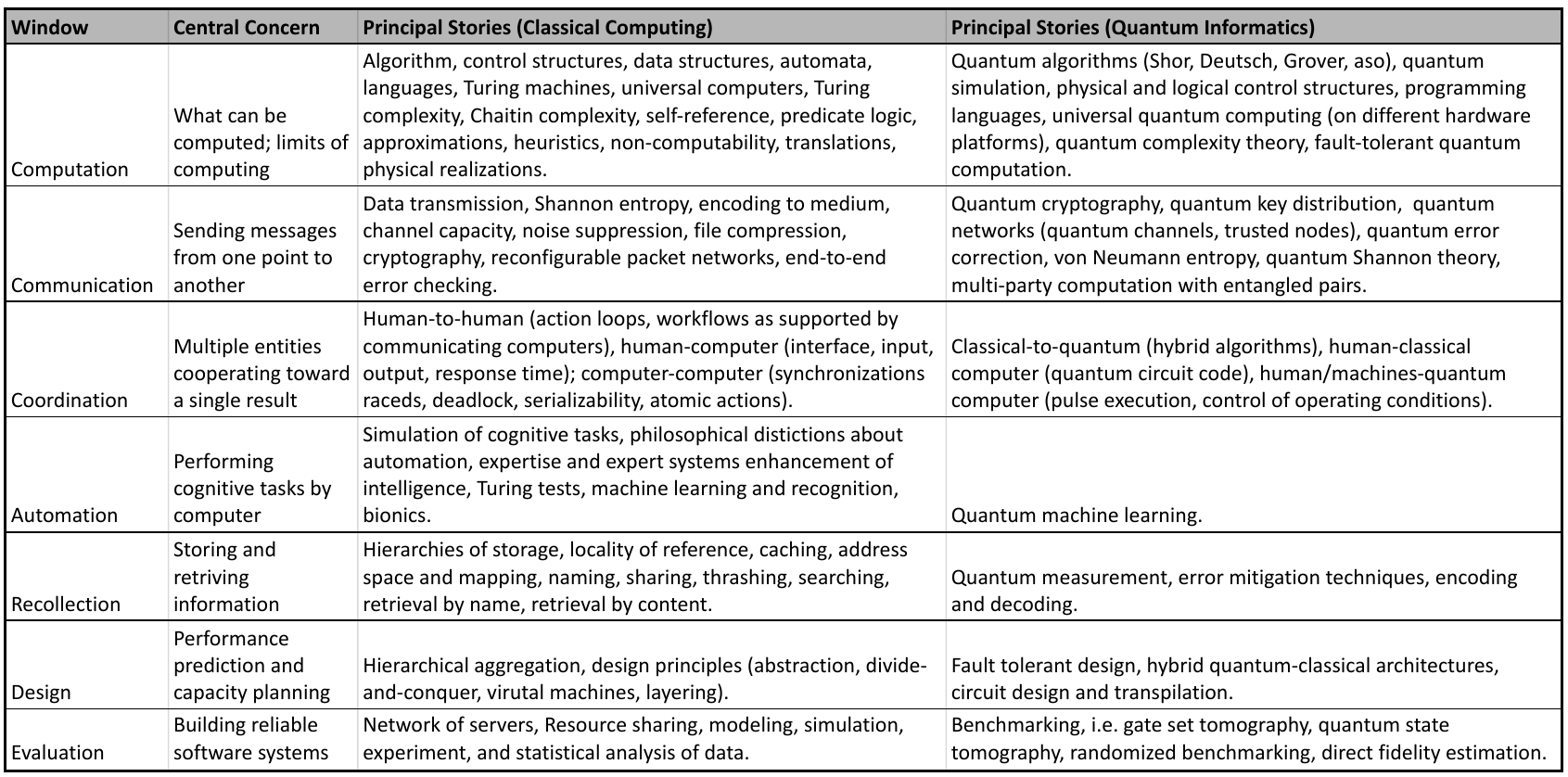}\end{table}

\section{Learning Standards for Quantum Informatics} \label{standards}

The K-12 Computer Science Framework \cite{computer_science_teacher_association_csta_nodate} is divided into practices and concepts that form the basis of our Quantum Informatics Standards. We left the practices unchanged because they are overarching practices that apply to computing in general as well as to its quantum applications. The standards were then formulated following the principles of the Standards Developer's Guide \cite{k12_computer_science_framework_guidance_nodate} highlighted in italics. Each standard was formulated by \textit{integrating one or more practices} with a quantum concept and one or more CS concepts. We formulated the standards with a high degree of \textit{rigor}, meaning that the standards should represent an appropriate cognitive and content demand for students. However, these are only theoretically formulated standards that need to be tested empirically. In light of future empirical evaluation, when unsure, we rather included a topic or chose a higher standard. For reasons of \textit{manageability}, we focused on fundamental concepts like quantum information and quantum algorithms, while left out topics such as quantum simulation and hardware for quantum information processing, as these were either less relevant for computer science standards or too complex for secondary schools. 
We wrote the standards to be \textit{specific} enough to convey and \textit{measure} the knowledge expected of students, and to be as jargon-free and accessible as possible for a topic as new and complex as quantum informatics (for the sake of clarity and consistency, we could not avoid using some technical terms in the standards definitions, but we have accompanied them with comprehensive explanations).  \textit{Equity and diversity} is a particularly important principle for our topic, since what we want to contribute to by teaching quantum informatics early is indeed a more diverse and inclusive science. In formulating the standards, we have been careful to ensure that they can be learned and demonstrated in a variety of ways, using different visualizations and representations, allowing learners to use and further develop individual learning strategies and approaches. However, this is an aspect that needs further and deeper attention in the future as more materials and approaches are developed. 
Last, the \textit{connection to other disciplines} is intrinsic to physics and mathematics, but has not been made explicit in this first formulation of the standards. 
Overall, it should be considered that these standards are a first theoretical formulation, and we wish to see them used, tested and critically evaluated in order to formulate more accurate, inclusive and adequate descriptions in the future. 

The resulting standards are summarized in Table \ref{tab:kompetenzmatrix}. For each quantum concept, we have indicated the corresponding CSTA concept. We did not establish a one-to-one correspondence with the CSTA concepts, as topics such as quantum error correction and quantum cryptography are very relevant within quantum informatics, while quantum networks and quantum internet are not yet developed enough to be suitable for secondary schools.
In the following, each quantum informatics content area is described in more detail, based largely on some of the authors' previous work \cite{paparo} and on the description of the key concepts in the literature (\eg \cite{alpert_key_nodate}, \cite{seegerer_quantum_2021}).

\textbf{Quantum information}: A qubit is the basic unit of quantum information. Qubits, like bits, always have a well-defined value when measured, but unlike bits, they also exhibit quantum mechanical properties such as superposition and entanglement. Superposition means that a qubit can be in a state that is a combination of 1 and 0, and only when it is measured will it be either 0 or 1 with a probability determined by its previous state of superposition. Qubits can also be entangled, that is, connected in such a way that none of the entangled qubits can be described independently of the others. If one qubit (in a maximally entangled pure two-qubit state) is measured, the state of the other entangled qubit is also instantaneously determined, no matter how far apart the qubits are. These special properties can provide computational advantages, since it is possible, for example, to influence the state of two or more different, physically distant qubits by a single operation. This understanding of qubits and the laws that they follow is the basis for further understanding of quantum informatics. In the classroom, students should learn what a qubit is and what makes it similar to bits and what makes it different. A deeper understanding of the special properties of qubits would allow students to grasp the possibilities as well as the limitations of quantum information and computation.  Students should also be able to read and evaluate the accuracy of a popular article on quantum computing. 

\textbf{Quantum error correction}:
 The current phase of quantum computer development is called the Noisy Intermediate-Scale Quantum (NISQ) era. These processors are very error-prone and are yet too small for implementing large-scale quantum error correction codes on them. Quantum error correction and fault-tolerant quantum computation are currently a very actively researched area and fundamental for the development of large scale quantum computers. Although classical error detection and correction requires students to apply important computer science concepts such as data decomposition and representation, it is not traditionally taught in school. By learning about quantum error correction, students will learn what error correction is, why it is relevant to computer science, and why classical error correction is not applicable to quantum computing. 

\textbf{Quantum computing systems}: Different physical methods of building quantum computers are currently being explored, such as superconducting qubits or ion traps. Just as there are different ways of developing quantum computers, there are also different languages for programming them. By using one of these platforms, students can experience the principles of quantum computing and interact with a quantum computer themselves. 

\textbf{Quantum algorithms}:
Calculations on quantum computers are described by quantum algorithms, and one model to describe them is that of quantum circuits. In a quantum circuit, the steps of the algorithm are described by quantum gates performed on one or more qubits and a measurement operation for readout.
So far, there are only a limited number of useful algorithms that could give quantum computers a significant speed advantage in basic computing problems such as factoring large numbers. However, for many other types of computation, there are no easy ways to implement them on a quantum computer, and there is no advantage to doing so. A current challenge in quantum computing is to develop efficient as well as useful algorithms for quantum computers. In fact, students should not be expected to develop new algorithms themselves. However, they can learn about the effects of gates on qubits and how they can be put together to create an algorithm. Students can replicate some well-known algorithms and reason about their logic and properties.

\textbf{Quantum cryptography}: Shor's algorithm showed that fully functional quantum computers (with a large enough number of qubits and a sufficiently small error rate) would be able to factor large numbers efficiently, threatening today's most widely used encryption methods. However, quantum effects can also make communication more secure based on physical principles.

\begin{landscape} 
\begin{table}[h]
  \centering
  \footnotesize
  \caption{Learning Standards for quantum informatics at secondary school level}
 \begin{tabular}{>{\columncolor{lightgray}}P{11em}>{\columncolor{white}}P{11em}>{\columncolor{lightgray}}P{27em}>{\columncolor{white}}P{10em}}
\hline
    \textbf{CSTA Concept} & \textbf{Quantum Concept} & \textbf{Standards} \newline{} By the end of grade 12 students should be able to: & \textbf{CSTA Practice} \\
    \hline\hline
    Data \& Analysis & Quantum Information & - point out the differences and similarities between qubits (quantum bits) and bits as units of information.
    \newline - explain which new possibilities of information processing arise with qubits (thanks to superposition and entanglement). & abstraction, computational problems \\
    
      & Quantum Error Correction & - explain why quantum error correction is fundamental in the development of quantum computers. \newline - describe the concept of fault tolerance. & computational problems\\
    \hline
    Computing Systems & Quantum Computing Systems & - describe the basics of quantum computer design and operation.
    \newline - use a formal language to interact with a quantum computer. & creating, testing and defining \\
    \hline
    Algorithms \& Programming & Quantum Algorithms & - explain how quantum algorithms can help solve important problems in computer science.
    \newline - rebuild given quantum algorithms and analyze their advantages over classical algorithms. & testing and defining, computational problems \\
    \hline
    Networks \& \newline{}the Internet & Quantum Cryptography & - describe why today's most commonly used encryption method is threatened by future quantum computing.
    \newline - test a quantum key distribution protocol (e.g., BB84)
    \newline - explain the advantages of quantum encryption and how it differs from other encryption methods. & collaborating, testing and refining \\
    \hline
    Impact of Computing & Impact of Quantum Informatics & - identify future implications between quantum informatics and its embedding in society.
    \newline - discuss choices, norms, and behaviors that are possible in response to the opportunities and risks of quantum computing. & communicating, inclusion \\
    \hline
  \end{tabular}
  \label{tab:kompetenzmatrix}
\end{table}
\end{landscape}

 Quantum cryptography takes advantage of the fact that due to the principles of quantum mechanics, it is impossible for a third party to eavesdrop on the system without disrupting it and thus (probably) being detected. Cryptography is fundamental to today's digital society, and many frameworks and curricula have recognized it \cite{lodi_cryptography_2022}. Students can learn about quantum cryptography in a course focused solely on cryptography, or while learning about quantum informatics. Students should learn why today's most widely used encryption method, RSA, is threatened by quantum computers. Also, by trying out a quantum key distribution protocol on their own, they can consider how it differs from classical cryptography and where its potential and risks lie.

\textbf{Impact of quantum informatics}:
If fully functional quantum computers are realized, they could crack the most commonly used encryption methods. At the same time, they could be used to perform process optimizations that could lead to significant efficiency gains and bring energy savings. In addition, quantum simulation has the potential to contribute to a better understanding of quantum mechanical systems, which could allow us, for example, to develop new drugs and materials.  Societal implications lend themselves well to discussion with students, and one could discuss, for example, how access to quantum computing, if limited to individual governments or a few private companies, could alter power relations in society (cf. \cite{de_wolf_potential_2017}).

\section{Exemplary Teaching approaches} \label{teachingexamp}
Lastly, we provide concrete examples of how the standards could be implemented in practice. We selected teaching approaches that fit with the newly defined Quantum Informatics Standards (selecting for content, age group and computer science applicability). In order to show how it is possible to develop different approaches to the Quantum Standards, appealing to younger students and different levels of knowledge and abstract thinking, we then classified the selected approaches based on Bruner’s modes of representation \cite{bruner}. These approaches are meant to be examples, and as such we did not undertake a systematic review of all existing approaches, as it was not the focus of this work.

\textbf{Action based approaches}: The role-playing approach to introduce to qubits and their properties developed by López-Incera and Dür \cite{lopez-incera_entangle_2019} is a good example for an action based introduction to \textit{quantum information}. The authors developed a role-playing game where some students are the qubits and others are the scientists. The qubits have to follow rules about how to position arms and legs and what to do when the scientists throw a ball at them (measure them), while the scientist have to figure out these rules. This way it is possible to explain superposition and entanglement, as well as the complexity of scientific hypotheses, in a tangible and playful way. The authors developed a similar role playing approach also for \textit{quantum cryptography} \cite{lopez-incera_encrypt_2020} and there are several other unplugged examples to teach to quantum cryptography. Perry \etal \cite{perry_quantum_2020} developed a pen\&paper way to experience the BB84 protocol, one of the main quantum key distribution protocols, which relies on qubit properties instead of mathematical complexity. Another promising enactive way to learn about quantum key distribution is to build and use the Qeygen machine \cite{tuftelakademie_quantum_nodate}. With this analog machine, students can exchange a key using the BB84 protocol, simulate an eavesdropper, and experience the possibilities and limitations of quantum key distribution.

\textbf{Image based approaches}: The states of a qubit are generally represented as vectors in a 2-dimensional complex space (Hilbert space), and most textbooks use the Bloch sphere (or a simplified unit circle) as a geometrical representation of qubits (\cite{billig_quantum_2018}, \cite{perry_quantum_2020}). However, since most high school students lack knowledge of complex numbers and linear algebra, this might not be the most accessible visualization. 
Often metaphors are used, such as flipping coins, balls of different colors, or a couple in love ordering wine in a restaurant.  Although the metaphors have the advantage of making a connection to the everyday life of the students, as also pointed out by Seegerer \etal \cite{seegerer_quantum_2021}, they run the risk of being taken too literally on the one hand, and of not making the peculiarity of quantum principles sufficiently obvious on the other. Therefore, also other approaches that are closer to quantum mechanical formalism have been developed, such as the QI4Q formalism \cite{rudolph_computing_nodate}. Here black and white marbles are used to represent qubits and the boxes the marbles pass through represent \textit{quantum gates}. This approach has the advantage of explaining all necessary quantum gates correctly and in an accessible, tangible way, without using any mathematics, but also the disadvantage of having to learn somewhat arbitrary rules and of being suitable only for relatively simple algorithms (cf. \cite{economou_teaching_2020}). Economou \etal \cite{economou_teaching_2020} showed how to use the QI4Q formalism to model the Deutsch algorithm, disguised as a game. This allows students not only to build a \textit{simple quantum algorithm} using the properties of quantum gates, but also to observe the advantages of quantum over classical information processing. As with classical algorithms, it is valuable for students to discuss them with pseudocode or different approaches before being confronted with a formal language.

\textbf{Language based approaches}: Many platforms offer free access (after registration) to their quantum computing power via the cloud, such as Qutech's Quantum Inspire \cite{Qutech} or IBM Quantum \cite{IBM}. This last one is widely used for educational purposes as it provides an attractive graphical interface where you can drag and drop to simulate (and execute) the effect of different gates on one or more qubits, as well as qiskit, a Python-based software development kit. Students can build their qubit circuits and run them on a real quantum computer, \textit{experiencing quantum computing} and its limitations, \ie they can compare the simulation with the real computer and understand the need for error mitigation as well as \textit{error correcting techniques}. The IBM website offers a comprehensive textbook with integrated exercises \cite{qiskit}, but it is aimed at university students with a high motivation and interest in mathematics. This could be reduced and simplified according to the students' knowledge, needs and the time available (as in \cite{introquantum}).
On a symbolic, language-based level, students can also discuss \textit{risks and potentials} in the development of quantum informatics. Although most learning resources mention this aspect, it is never the main focus of the teaching material on quantum informatics. Interesting approaches to thematize this can however be borrowed by other disciplines such as future studies \cite{future} and integrated in a quantum informatics curriculum.

The presented approaches show how an action based access to quantum information and quantum cryptography is possible for younger students (e.g. lower secondary school), while more complex and abstract topics such as quantum computer systems, quantum algorithms, and the implications of quantum informatics might be more suitable for higher secondary school. In general, we have provided one or more teaching examples for the defined quantum concepts, although not all defined learning outcomes, such as the ability to describe the concept of fault tolerance or the discussion of responses to the opportunities and risks of quantum computing, are directly addressed by existing approaches. We are convinced that it is possible to teach all the defined standards in a way that is appropriate for school, and intend to show this in the future.

\section{Conclusion}
With this paper, we offer a contribution to the introduction of quantum informatics in schools: Its technologies and practices have been mapped and viewed from the perspective of the Great Principles framework, and a first proposal for learning standards at secondary school level, have been presented.
It is important to note that this is only preliminary theoretical work and that the standards need to be tested and evaluated empirically. We are in the process of developing teaching materials and approaches based on the proposed standards and the analysis of existing teaching approaches, and hope that others will do the same. It is our goal that the standards and analysis proposed here will be useful to computer science teachers and educators who are designing new materials, planning lessons, or seeking a first orientation for teaching quantum informatics.

%
%
%
\bibliographystyle{splncs04}
\bibliography{ISEEP}

\end{document}